\documentstyle[12pt,aasms4]{article}
\lefthead{FRYER \& KALOGERA}
\righthead{DOUBLE NEUTRON STARS AND KICK VELOCITIES}
\begin{document}
%\begin{center}
%To appear in {\em The Astrophysical Journal}
%\end{center}
\vspace{1.cm}
\title{Double Neutron Star Systems and Natal Neutron Star Kicks}
\author{Chris Fryer}
\affil{Lick Observatory, University of California Observatories, \\ 
Santa Cruz, CA 95064 \\ cfryer@ucolick.org}
\authoremail{cfryer@ucolick.org} 
\author{Vassiliki Kalogera} 
\affil{Astronomy Department, University of Illinois at
Urbana-Champaign, \\ 1002 West Green St., Urbana, IL 61801 \\
vicky@astro.uiuc.edu}
\authoremail{vicky@astro.uiuc.edu} 

\begin{abstract} 

We study the four double neutron star systems found in the Galactic disk
in terms of the orbital characteristics of their immediate progenitors and
the natal kicks imparted to neutron stars. Analysis of the effect of the
second supernova explosion on the orbital dynamics, combined with 
recent results from simulations of rapid accretion onto neutron stars
lead us to conclude that the observed systems could not have been formed
had the explosion been symmetric. Their formation becomes possible if
kicks are imparted to the radio-pulsar companions at birth. We identify
the constraints imposed on the immediate progenitors of the observed
double neutron stars and calculate the ranges within which their binary
characteristics (orbital separations and masses of the exploding stars)
are restricted.  We also study the dependence of these limits on the
magnitude of the kick velocity and the time elapsed since the second
explosion. For each of the double neutron stars, we derive a minimum kick
magnitude required for their formation, and for the two systems in close
orbits ($\lesssim 10$\,R$_\odot$), we find it to exceed 200\,km/s.  Lower
limits are also set to the center-of-mass velocities of double neutron
stars, and we find them to be consistent with the current proper motion
observations. 

\end{abstract}


\section{Introduction}

The Double Neutron Star (DNS) system containing PSR 1913+16, discovered
in 1975 by Hulse \& Taylor\markcite{Hul75}, was the first binary pulsar
for which observational data strongly suggested that both components were
compact objects.  Following PSR 1913+16\footnote{Hereafter, we will refer
to the double neutron star systems using the designation of their radio
pulsar component.}, additional DNS candidate systems have been
discovered: PSR 2303+46 (Stokes, Taylor, \& Dewey\markcite{Sto85} 1985),
PSR 2127+11C (Anderson et al.\markcite{And90} 1990), PSR 1534+12
(Wolszczan\markcite{Wol91} 1991), and PSR 1518+4904 (Nice, Sayer, \&
Taylor\markcite{Nic96} 1996). The high orbital eccentricities of these
systems separate them from all other binary radio pulsars, which are
found in circular orbits (e.g., van den Heuvel\markcite{Heu95} 1995).
Measured and derived orbital and pulsar properties of these five systems
are given in Table 1. 

One of the two members in each binary is a neutron star since it is
detected as a radio pulsar.  Using a variety of tests, the nature of the
companion to the pulsar has been investigated in detail for PSR 1913+16
(Taylor \& Weisberg\markcite{Tay82}\markcite{Tay89} 1982, 1989)  and to a
lesser extent for the other systems.  Extended companions are excluded
based on their radial size, their implied periastron advances, and their
apparent visual magnitudes, while helium stars of the same mass are even
brighter (Crane, Nelson, \& Tyson\markcite{Cra79} 1979; van
Kerkwijk\markcite{Ker96} 1996; Webbink\markcite{Web96} 1996). For three 
of the systems, white dwarf companions of various temperatures can
potentially be ruled out using deep optical observations (see Table 2).
For the two most distant systems, PSR 1913+16 and 2127+11C, there is no
current observational test that can rule out white dwarf companions. 
However, the evolutionary history and high orbital eccentricities of
these systems require that the pulsar companions were formed in supernova
explosions that occurred after the pulsar spin-up phase (Srinivasan \&
van den Heuvel\markcite{Sri82} 1982). This provides indirect evidence
that all of the pulsar companions are either neutron stars or black
holes. 

The evolution of double neutron stars after the second supernova explosion is
governed only by angular momentum losses due to emission of gravitational
waves, and is not complicated by effects of binary-star evolution (e.g.,
nuclear evolution, mass transfer, stellar winds). These systems, therefore,
constitute an ideal diagnostic for neutron star kicks.  Natal kicks are
thought to be imparted to neutron stars at the time of their formation as a
result of some asymmetry in the supernova explosion. Evidence for the
existence of such kicks originates in observations of radio pulsar velocities
(Caraveo\markcite{Car93} 1993;  Harrison, Lyne, \& Anderson\markcite{Har93}
1993; Frail, Goss, \& Whiteoak\markcite{Fra94} 1994; Lyne \&
Lorimer\markcite{Lyn94} 1994), O,B runaway stars (Stone\markcite{Sto91}
1991), and X-ray binaries (van den Heuvel \& Rappaport\markcite{vdH87} 1987;
Kaspi et al.\markcite{Kas96} 1996), and estimates of their average magnitudes
vary from $\sim 100$ to $\sim 500$\,km/s (Dewey \& Cordes\markcite{Dew87}
1987; Bailes\markcite{Ba89} 1989; Fryer, Burrows, \& Benz\markcite{Fry97}
1997).

In this paper, we thoroughly investigate the four double neutron star systems
found in the Galactic field to identify the allowed binary characteristics of
their immediate progenitors and to constrain the nature of the natal kick
required for their formation.  In \S\,2, we examine the possibility that DNSs
are formed via symmetric supernova explosions. Based on arguments related to
orbital dynamics and, in contrast to previous work, accounting for orbital
evolution due to gravitational radiation, we find that the first neutron star
in the observed systems always lies within the envelopes of their exploding
companions.  In the past, similar conclusions were reached for some of the
systems (Burrows \& Woosley\markcite{Bur86} 1986; Yamaoka, Shigeyama, \&
Nomoto\markcite{Y93} 1993), although the fate of the engulfed neutron stars
was not addressed in any detail. We use recent results on neutron-star
common-envelope phases, and conclude that the second supernova explosions in
DNS systems could not have been symmetric. In \S\,3, we account for the
effects of kicks imparted to nascent neutron stars and find that the observed
systems can then be reproduced. Previous studies (Burrows \& Woosley 1986;
Wijers, van Paradijs, \& van den Heuvel\markcite{Wij92} 1992; Yamaoka,
Shigeyama, \& Nomoto\markcite{Y93} 1993) have considered kicks of specific
magnitude or direction or have indicated plausible values of progenitor
parameters and kick magnitudes.  Here, we are able to (i) identify the ranges
of values within which the properties of the immediate progenitors of the
observed DNSs are restricted, and (ii) derive an absolute lower limit to the
magnitude of the kick velocity needed for their formation. We conclude with a
discussion of the implications of our results for the intrinsic kick velocity
distribution and the center-of-mass velocities of double neutron stars.

\section{Symmetric Explosions}

The existence of double neutron stars has instigated studies of their origin
and a variety of evolutionary sequences have been proposed for their
formation (Flannery \& van den Heuvel\markcite{Fla75} 1975;
Webbink\markcite{Web75} 1975; Smarr \& Blandford\markcite{Sma76} 1976; 
Srinivasan \& van den Heuvel 1982;  van den Heuvel, Kaper, \&
Ruymaekers\markcite{van94} 1994; Brown\markcite{Bro95} 1995; Terman \&
Taam\markcite{Ter95} 1995). Regardless of the details of prior evolution, all
mechanisms converge to the same configuration before the formation of the
second neutron star: a binary consisting of a neutron star and a helium star
in a circular orbit.  The helium star then collapses to form the companion to
the observed radio pulsar.  A possible exception is PSR 2127+11C, which is
found in the dense stellar environment of a globular cluster and may have
formed via dynamical interactions (capture or exchange: Anderson et al. 
1990). Therefore, we restrict our analysis to the four DNS systems found in
the Galactic field.

In this section, we assume that the supernova explosion of the helium
star is symmetric, its only effect being an instantaneous mass loss from
the system. Since the companion of the exploding star is a neutron star,
we may neglect any interaction between the radio pulsar and the supernova
ejecta. Combining the results for the pre-SN configuration with recent
developments on common envelope evolution of inspiraling neutron stars,
we conclude that the four double neutron star systems found in the
Galactic field could not have been formed if supernova explosions were
symmetric. 

\subsection{Orbital Dynamics}

In order to obtain the post-SN parameters for each of the known systems,
we must account for orbital evolution due to gravitational wave emission
(Wijers, van Paradijs, \& van den Heuvel\markcite{Wij92} 1992 recognized
the distinction between present and post-SN characteristics, but did not
include this effect in their analysis). For each system, the extent of
orbital evolution depends on the time, $T_{\rm SN}$, elapsed since the
supernova explosion.  Since the radio pulsar has been recycled prior to
the supernova explosion, its age sets an upper to the value of $T_{\rm
SN}$. The difference between the two times is of the order of the
lifetime of the helium star ($\sim 10^6$\,yr; see Habets\markcite{Hab85}
1985). The ``characteristic age'' of the pulsar defined as $\tau_c\equiv
P_s/2\dot{P}_s$ (where $P_s$ and $\dot{P}_s$ are the pulsar spin period
and its derivative) is often used as a measure of the pulsar age.
However, this estimate is strongly model dependent (assumptions about the
form of the braking law, the pulsar magnetic field and the initial spin
period), and hence the upper limit on $T_{\rm SN}$, may be considerably
different than the value of $\tau_c$ (e.g., Lyne \&
Graham-Smith\markcite{Lyn90} 1990). Therefore, in our calculations, we
treat $T_{\rm SN}$ as a free parameter ranging from $0.1 \tau_c$ to $10
\tau_c$.  We use the expressions derived by Junker \&
Schaefer\markcite{Jun92} (1992)  for orbital evolution due to
gravitational radiation to integrate backwards in time, and calculate the
post-SN orbital separations, $A$, and eccentricities, $e$, from the
observed values ($A_{\rm cur}, e_{\rm cur}$).  This effect is strongest
for the two systems with short orbital periods, PSR 1913+16 and PSR
1534+12 (see Figure 1).

For a given set of post-SN characteristics, we can calculate the
corresponding parameters just prior to the explosion.  We denote these
parameters with a subscript ``$o$''. The radio pulsar mass, $M_p$,
remains the same before and after the explosion. 

\subsubsection{Circular Pre-SN Orbits}

In the case of circular pre-SN orbits, the orbital separation, $A_o$, and
mass, $M_o$, of the exploding star are given by (see also Blaauw 1961; 
Boersma\markcite{Boe61} 1961):
\begin{equation} 
\frac{A_o}{A}~=~\frac{M_p\,+\,2\,M_c\,-\,M_o}
{M_p\,+\,M_c}\mbox{,} 
\end{equation} 
and
\begin{equation}
M_o~=~e\,(M_p\,+\,M_c)\,+\,M_c.  
\end{equation} 

According to the current formation mechanisms, the exploding star is an 
helium star. During the course of their
evolution, helium stars less massive than $\sim$3.5\,M$_\odot$ expand
significantly. We use a fit for the maximum helium-star radius,
$R_{\rm He,max}$, as a function of its mass,
$M_{\rm He}$,
(Kalogera \& Webbink\markcite{KW96} 1996) based on helium star models
calculated by Habets (1985) (which are consistent with
the models of Woosley, Langer, \& Weaver\markcite{Woo95} 1995):   
\begin{equation} 
\mbox{log} R_{\rm He,max}=
\left\{ \begin{array}{ll} 3.0965-2.013 \, \mbox{log} \, M_{\rm He} &
M_{\rm He} \leq 2.5
\mbox{M}_{\odot} \\ 0.0557 \,(\mbox{log} \, M_{\rm He} - 0.172)^{-2.5}
& M_{\rm He} >
2.5
\mbox{M}_{\odot}
.  \end{array} \right. 
\end{equation} 

Using equations (1), (2), and (3) we calculate the pre-SN separations,
$A_o$, progenitor masses of the pulsar companions, $M_o$, and the ratios
of orbital separations over helium-star maximum radii,
$A_o/R_{\rm He,max}$, as functions of $T_{\rm SN}$. The
results for PSR 1913+16 and PSR 1534+12 are shown in Figure 2, while the
ranges of values for all systems are given in Table 3.  It is evident
that for all observed DNSs and for a wide range of $T_{\rm SN}$, the
stellar radius of the helium star exceeds the orbital separation. In
other words, if the explosion were symmetric, we find that in the
progenitors of {\em all} the observed systems, the radio pulsar would lie
within the envelope of its companion at the time of the supernova
explosion (see also Burrows \& Woosley 1986; Yamaoka et al. 1993).

\subsubsection{Eccentric Pre-SN Orbits}

All current formation scenarios for double neutron star systems lead to
circularization of the pre-SN orbits as the first neutron star gets
recycled.  However, the pulsars may not be recycled or there may be a way
to recycle the pulsar without circularizing the orbit of the system. 
Hence, we examine our conclusion that the pulsar would be in a common
envelope phase at the time of the explosion, in the case of eccentric
pre-SN orbits. 

In this general case, the relations between the pre- and post-SN
parameters have been derived by Hills\markcite{Hil83} (1983): 
\begin{equation}
\frac{A}{A_o}~=~\frac{(M_o+M_p)\,-\,(M_o-M_c)}
{(M_o+M_p)\,-2\,(A_o/r)\,(M_o-M_c)}\mbox{,}
\end{equation}
and
\begin{equation}
e~=~\left \{ 1 - (1-e_o^2)
\left(1\,-\,2\,\frac{A_o}{r}\,\frac{M_o-M_c}
{M_o+M_p}\right)\,\left(1\,-\,\frac{M_o-M_c}{M_o+M_p}\right)^{-2}\right
\}^{1/2} \mbox{,}
\end{equation}
where $r$ is the distance between the two stars with masses $M_o$ and
$M_p$, at the time of the supernova explosion.  There are four unknown
pre-SN parameters: the orbital separation, $A_o$, the helium-star mass,
$M_o$, the pre-SN eccentricity, $e_o$, and the distance, $r$, between the
two stars at the time of the explosion.  Since we only have two
equations, we treat two of the above quantities, $e_o$ and $M_o$, as
parameters and solve for $A_o$ and $r$.  Note that $r$ can acquire values
only within a restricted range, between periastron and apastron of the
pre-SN orbit. 

The ratio of the periastron distance to the maximum radius for the helium
star as a function of $M_o$ and for a complete range of pre-SN eccentricities
is shown in Figure 3 for all four of the observed systems (see also Table 3).
We see that although for PSR 2303+46 and 1518+4904 the ratio
$A_o(1-e_o)/R_{\rm He,max}$ might exceed unity, for PSR 1913+16
and 1534+12 the pulsar would lie well within the envelope of the helium star.
Therefore, if the supernova explosions were symmetric, the pulsars in at
least two of the observed DNS systems would experience common envelope
evolution, even in the extreme case of eccentric pre-SN orbits.

\subsection{Neutron Stars in Common-Envelope Phases}

For a symmetric explosion, the immediate progenitors of double neutron
star systems must evolve through a common-envelope phase. In this
section, we use recent results on rapid accretion onto neutron stars to
examine their fate in the envelopes of evolved helium stars. We conclude
that survival of neutron stars through such phases is not possible.

Accretion of matter onto the neutron star in common envelope phases has
been recently studied by various investigators
(Chevalier\markcite{Che93}\markcite{Che96} 1993, 1996; Brown 1995; Fryer,
Benz, \& Herant\markcite{Fry96} 1996).  In common envelope situations,
the Bondi-Hoyle accretion rate (Bondi\markcite{Bon52} 1952)  typically
exceeds $\sim 10^{-4}$\,M$_{\odot} {\rm yr}^{-1}$, which is much greater
than the photon Eddington limit for neutron stars.  For such high
accretion rates, neutrino emission dominates the cooling processes. 
Accretion is not limited by the photon Eddington limit, but instead by
the corresponding limit for neutrinos, which is 20 orders of magnitude
greater than the photon Eddington limit.  For the calculated Bondi-Hoyle
rates, neutrino emission carries away the energy released by accretion, a
steady state accretion flow at high rates is achieved, and the pulsar
collapses into a black hole on time scales of hours or months. 

Depending on the conditions in the common envelope, other physical
processes can facilitate the survival of the neutron star.  Chevalier
(1996) showed that the angular momentum of the accreted material may
prevent rapid accretion and hence collapse of the neutron star into a
black hole. Also, Fryer et al. (1996) found that the neutrinos emitted
close to the neutron star may be able to heat a region of the infalling
atmosphere, and ultimately drive a small scale explosion which temporarily
halts mass accretion. A series of such explosions may also prohibit the
collapse of the neutron star into a black hole if a sizable amount of the
atmosphere can be blown off of the neutron star instead of being accreted. 
Both of these effects depend sensitively on the Bondi-Hoyle infall rate
onto the neutron star and they become more important as the infall rate
decreases. The envelopes of hydrogen-rich giant stars (with masses
$\sim 10$\,M$_\odot$) are less dense (by at least one order of magnitude)
than those of helium-rich giants and have correspondingly lower infall
rates.  Although, in helium giants, the infall rates are too high for
angular momentum to limit the flow, it may well limit the accretion for
some hydrogen giants (see Table 1 in Fryer et al. 1996). 

In this section, we calculate the time, $T_{\rm coll}$, needed for a
neutron star to collapse into a black hole (for zero and maximum angular
momentum of the material), and compare it to the evolutionary time,
$T_{\rm ev}$, from maximum radial extent of the helium star to its
explosion.  Only if $T_{\rm ev}$ is shorter than $T_{\rm coll}$ the
helium star will explode in a supernova before the neutron star collapses
into a black hole. 

For each of the observed systems and under the assumption of symmetric
supernova explosions, we have derived the mass of the progenitor of the
second neutron star as well as the pre-SN orbital separation, that is the
position of the pulsar in the envelope of its companion at the time the
core of the helium star collapses.  Using helium star models by
Woosley\markcite{Woo96} (1996)  we calculate the Bondi-Hoyle accretion
rates, $\dot{M}_{\rm B}$, appropriate for the physical conditions in the
helium envelopes (see Table 4). We have assumed that the neutron star
velocity in the envelope is equal to the pre-SN relative orbital
velocity. These Bondi-Hoyle rates are too high for the explosions found
by Fryer et al.  (1996) to occur. In the extreme case that angular
momentum has no effect on accretion, matter is accreted by the neutron
star at the Bondi-Hoyle rate. Thus, we can calculate the corresponding
time scale for collapse, $T_{\rm coll}$, into a black hole, assuming that
the amount of mass required for collapse is $\sim 1$\,M$_\odot$ (see
Table 4). 

Matter accreted onto the neutron star is expected to have angular
momentum due to density and velocity gradients in the envelope. As a
result, infall of material is halted at all polar angles except in a cone
around the poles defined by the angular momentum symmetry axis. For the
extreme case of negligible angular momentum transport, we derive a lower
bound to the accretion rate onto the neutron star. For a given value of
the specific angular momentum, $j$, there is a critical polar angle,
$\theta_c$, within which angular momentum does not affect the inflow
appreciably.  This angle is set by the balance between the centrifugal
and gravitational forces at the neutron star radius (in the Newtonian
limit):
\begin{equation}
\frac{j^2\,\mbox{sin}^2\theta_c}{R_{\rm NS}^3}~=~\frac{G\,M_{\rm NS}}
{R_{\rm NS}^2}
\mbox{,} 
\end{equation} 
where $M_{\rm NS}$ and $R_{\rm NS}$ are the gravitational
mass and radius of the neutron star, assumed here to be equal to
$1.4$\,M$_\odot$ and $10^6$\,cm, respectively. For the specific angular
momentum
we use the analytical estimates by Ruffert \& Anzer\markcite{Ruf95} (1995): 
\begin{equation} 
j =
\frac{1}{4}(6\epsilon_v-\epsilon_{\rho})\,V\,R_{\rm B}\mbox{,}
\end{equation}
where $V$ is the velocity of the neutron star, $R_{\rm B}$ is the
Bondi-Hoyle accretion radius (Bondi 1952), and $\epsilon_{v}$,
$\epsilon_{\rho}$ are the velocity and density gradient parameters, as
defined in Ruffert \& Anzer (1995). We note that their numerical
calculations indicate that the above expression (eq. [7]) overestimates
the amount of angular momentum accreted onto the neutron star, and
therefore the corresponding mass accretion rate we derive is an
underestimate. Using the helium-star models by Woosley (1996) we
calculate $j$ and $\theta_c$. For accretion restricted within the polar
regions, we obtain an extreme lower limit to the accretion rate onto the
neutron star and a corresponding upper limit to the characteristic time
scale for collapse, $T_{\rm coll}$.

The neutron star will avoid the collapse into a black hole only if
$T_{\rm coll}$ is longer than the evolutionary time interval,
$T_{\rm ev}$, between the time the helium star acquires its maximum
radial extent and the time it explodes as a supernova. For all of the
observed systems (with the possible exception of PSR 1518+4904), the
collapse time scale is many orders of magnitude {\em shorter} than the
evolution time scale, which clearly indicates that the pulsar becomes a
black hole long before the SN explosion.

In view of the above analysis, we conclude that the pulsar cannot survive
a common-envelope phase with its helium star companion.  Nevertheless,
such a phase could not be avoided if the explosion were symmetric, and
hence would not lead to the formation of the double neutron star systems.

\section{Asymmetric Explosions}

We have already shown that if supernovae explosions were symmetric, the
first-born neutron stars would have collapsed into black holes during a
common-envelope phase with their helium-star companions, thus aborting
DNS formation.  In this section, we find that asymmetric explosions,
during which kicks are imparted to nascent neutron stars, allow the
formation of the observed double neutron star systems (see also Yamaoka
et al. 1993). The pre-SN parameters of their progenitors are still
restricted within relatively narrow ranges, which depend on the magnitude
of the kick velocity, the time elapsed since the SN explosion, and the
eccentricity of the pre-SN orbits.  For each of the systems we are able
to evaluate the minimum kick magnitude required for its formation. 

\subsection{Circular Pre-SN Orbits}

First, we study the case of circular pre-SN orbits, consistent with all
current DNS formation mechanisms. Using the orbital energy and angular
momentum equations for eccentric orbits, and for a kick velocity of
specific magnitude and direction, expressions for the pre-SN orbital
separation, $A_o$, and mass of the helium star, $M_o$, as a function of
the post-SN binary characteristics can be derived (e.g., Hills 1983; 
Kalogera\markcite{Kal96} 1996):
\begin{equation}
V_k^2+V_r^2+2V_kV_r\mbox{cos}\theta~=~G\,(M_p+M_c)\,
\left(\frac{2}{A_o}-\frac{1}{A}\right)\mbox{,} 
\end{equation}
\begin{equation} 
A_o^2\,\left[V_k^2\mbox{sin}^2\theta\mbox{cos}^2\phi+
\left(V_k\mbox{cos}\theta+V_r\right)^2\right]~=~G\,(M_p+M_c)\,A\,
(1-e^2)\mbox{,} 
\end{equation} 
where $V_r$ is the relative orbital
velocity before the explosion, 
\begin{equation}
V_r~=~\left[\frac{G(M_o+M_p)}{A_o}\right]^{1/2}\mbox{,} 
\end{equation}
$V_k$ is the magnitude of the kick, $\theta$ is the polar angle of the
kick with respect to the pre-SN orbital velocity of the helium star
relative to the pulsar, and $\phi$ is the corresponding azimuthal angle,
so that $\phi=0$ represents a plane perpendicular to $A_o$. For each of
the observed DNS systems, the post-SN parameters are known. First, we
calculate the orbital evolution due to gravitational radiation assuming
$T_{\rm SN}=\tau_c$ (Table 1), and discuss how variations in
$T_{\rm SN}$ affect our results at the end of this section.  For a
specified kick magnitude in any direction, we calculate the two unknown
pre-SN parameters, $A_o$ and $M_o$.

The region of allowed pairs of values is restricted by a set of
constraints that DNS progenitors must satisfy. The first of the
constraints arises from the fact that the post-SN orbit must include the
position of the two stars at the time of the explosion (e.g., Flannery \&
van den Heuvel 1975; Yamaoka et al. 1993; Kalogera 1996), hence the
pre-SN separation must be within the bounds set by the periastron and
apastron distances in the post-SN orbit
\begin{equation} 
A \, (1-e) \leq A_o \leq A\,(1+e).  
\end{equation}
These two limits follow from equations (8) and (9) with the constraint
that $\cos^2\phi\leq 1$, where the equality corresponds to kicks
restricted to the plane perpendicular to $A_o$. 

A second constraint is set by demanding that $\cos^2\phi$ remains always
positive. Using equations (8) and (9) we find that for a specific pre-SN
separation the mass of the helium star is restricted to:  
\begin{equation}
M_o \leq \frac{2 \, k^2 (M_p+M_c)}{2 \, \alpha 
\left[2 \, \alpha^2(1-e^2)-k \right] - 4 \alpha^2 (1-e^2)^{1/2} 
\left[ \alpha^2 \, (1-e^2) - k \right]^{1/2}} - M_p \mbox{, }
\end{equation}
where 
\begin{equation}
k=2 \, \alpha - \left[ \frac{V_k^2 A}{G(M_p+M_c)} + 1 \right]
\end{equation} 
and $\alpha = A/A_o$.  The extreme case of $\cos^2\phi=0$ corresponds to
kicks restricted to the plane of the orbit before the explosion.

The final constraint arises from the condition that the pulsar must avoid
a common envelope phase with the helium star. The ranges of allowed
values for $A_o$ and $M_o$ are then restricted by the condition that the
helium star must fit within its Roche lobe when its radius is at a
maximum (eq. [3]).  At a given value of $A_o$, a lower limit is derived
for the mass of the helium star, since the maximum radii decrease with
increasing helium-star mass.

The limits imposed on pre-SN orbital separations, $A_o$, and
helium-star masses, $M_o$, are shown in Figures 4 and 5 for the two
observed systems for which the individual masses are accurately known (PSR
1913+16 and 1534+12). For the close DNS systems, the value of $A_o$ is
restricted in the range $\sim 3-4.5$\, R$_{\odot}$ and that of $M_o$ must
exceed $\sim 4$\,M$_{\odot}$, whereas, for the two wide systems (PSR
2303+46 and 1518+4904), it is $A_o \sim 20-50$\,R$_{\odot}$ and $M_o
\gtrsim 3$\,M$_{\odot}$. The solid thick vertical line corresponds to
the geometrical constraint (eq.\,[11]) and sets an upper limit on the
pre-SN orbital separation. Although there also exists a lower limit to
$A_o$, its value is too low to restrict the parameter space.  The thin
lines correspond to kicks imparted in the pre-SN orbital plane
($\cos^2\phi=0$), and thus to maximum helium star masses (eq.\,[12]); of
the three limiting constraints this is the only one that depends on the
assumed kick magnitude.  As the magnitude of the kick velocity increases,
the relative orbital velocity of allowed DNS progenitors also increases,
and systems with shorter separations and more massive helium stars are
included in the progenitor parameter space (see Figure 4).  For the two
close systems, PSR 1913+16 and 1534+12, and for a DNS age equal to
$\tau_c$, the pulsar companions must have been imparted a kick of at least
260 and 220\,km/s, respectively. For the wider systems, PSR 2303+46 and
1518+4904, the minimum kick magnitudes are lower, 70 and 50\,km/s,
respectively.

The parameter space allowed to DNS progenitors depends also on the time,
$T_{\rm SN}$, elapsed since the supernova explosion, during which
significant orbital evolution due to gravitational wave emission possibly
occurs. The limits on the ranges of pre-SN parameters are altered with
the exception of the limiting curve related to the size of evolved helium
stars (thick solid lines in Figures 4 and 5). As the value of
$T_{\rm SN}$ increases, post-SN separations and eccentricities
increase too, and hence the maximum allowed pre-SN separations are
shifted to higher values (eq.\,[11]). The consequent decrease in relative
orbital velocity before the explosion also results in a decrease of the
minimum kick magnitude required for the formation of the observed
systems.  For $T_{\rm SN}=10\,\tau_c$ (or $T_{\rm SN}=0$) the
minimum kick magnitude required decreases to 170\,km/s and 190\,km/s
(increases to 290\,km/s and 225\,km/s) for PSR 1913+16 and 1534+12,
respectively. For the two wide systems (PSR 2303+46 and 1518+4904)  the
effect of orbital evolution due to gravitational wave emission is
negligible, and therefore the minimum kick magnitudes remain unaffected. 

\subsection{Eccentric Orbits}

Although none of the current formation mechanisms requires an eccentric
orbit for the pulsar-helium star binary, as in the case of symmetric
explosions, we explore such a possibility and its effect on the pre-SN
binary parameters. In the case of eccentric orbits, the pre-SN parameter
space becomes four-dimensional: the eccentricity, $e_o$, and the position
along the orbits at the time of the explosion expressed by the eccentric
anomaly, $E_o$, are added to the orbital separation, $A_o$, and
helium-star mass, $M_o$. We sample the complete range of values for both
$e_o$ (0 to 1) and $E_o$ (0 to $\pi$), and calculate the outer envelope
of the limits imposed on pre-SN separation and mass of the exploding
star, based on the three constraints discussed above. These limits are
shown in Figure 6 for PSR 1913+16 and PSR 1534+12 for two different
values of the kick magnitude and for post-SN parameters corresponding to
$T_{\rm SN}=\tau_c$. The area of the parameter space allowed to DNS
progenitors is enclosed by the limiting curves. The lower limit on $M_o$
represents the constraint that the helium star at its maximum extent must
fit within the orbit. The upper limit corresponds to kicks restricted to
the orbital plane ($\cos\phi^2=0$). The condition $\cos\phi^2=1$ is no
longer restrictive for the progenitor parameters.  Since the explosion is
instantaneous, the distance, $r$, between the two binary members before
and after the supernova remains unchanged: 
\begin{equation}
r~=~A_o(1-e_o\,\mbox{cos}E_o)~=~A(1-e\,\mbox{cos}E)\mbox{.}
\end{equation}
In the limit of nearly parabolic pre-SN orbits and for $E_o=0$, no upper
limit can be set to the orbital separation, $A_o$. As the magnitude of
the kick increases, wider and more eccentric progenitor orbits and more
massive helium stars are allowed. However, we note that helium stars are
known to lose mass in strong stellar winds and thus their masses at
collapse are expected to be relatively small ($\lesssim 5$\,M$_\odot$)
(Woosley, Langer, \& Weaver 1995). Despite the expansion of the parameter
space allowed to DNS progenitors, the minimum kick magnitude required to
form the observed systems remains practically unaltered. 

\section{Discussion}

A careful account of the change in orbital characteristics of binary
systems experiencing symmetric supernova explosions, of the sizes of
helium stars approaching collapse into a neutron star, and of recent
results concerning the fate of neutron stars within the envelopes of
helium stars, led us to conclude that the observed double neutron star
systems in the galactic disk could not have been formed if the SN
explosions forming the pulsar companions were symmetric. Natal kicks
imparted to neutron stars are required to explain the observed parameters
of DNS systems, even if they have experienced significant orbital
evolution after their formation due to gravitational wave emission.  In
fact, only when kicks have magnitudes exceeding some minimum value can the
progenitor orbits be sufficiently wide to accommodate evolved helium
stars, and yet produce the small separations measured in these systems.
For the observed double neutron stars in close orbits, the minimum kick
velocities required are in excess of 200\,km/s.

The analysis presented here has enabled us to constrain the orbital
parameters of the immediate progenitors of the observed DNS systems. 
However, obtaining the most probable of the parameters within the allowed
ranges requires modeling of the evolutionary history of the DNS
progenitors up to the formation of the pulsar/helium-star binaries. We
have used the population synthesis code (Monte Carlo)  described in Fryer
et al. (1997) to obtain an estimate of the most probable kick velocities,
and hence progenitor parameters, for the observed systems. The standard
DNS formation mechanism (Srinivasan \& van den Heuvel 1982) is modeled,
including the constraint that the helium stars must be accommodated in
their orbits prior to collapse. The formation rate of DNS systems with
separations $\lesssim 5 {\rm R_{\odot}}$ peaks at $200$\,km/s
($5\times10^{-8}$\,yr$^{-1}$), and falls off sharply beyond $300$\,km/s.
Although higher kicks are allowed by our analysis, they disrupt most of
the binaries and do not contribute significantly to DNS formation.  For
double neutron star systems with separations $\lesssim 50$ \,R$_{\odot}$,
the formation rate ($\sim 10^{-6}$\,yr$^{-1}$) peaks at kicks lower than
100\,km/s, and is consistent with the rate inferred observationally (van
den Heuvel 1995).  These results indicate that kick velocities close to
the derived lower limits are favored. Consequently, progenitor orbital
separations and masses of $\sim 4.5$\,R$_\odot$ and $\sim 4.5$\,M$_\odot$
are favored for the close systems (PSR 1913+16 and 1534+12).  Separations
of $\sim 50$\,R$_\odot$ and $\sim 30$\,R$_\odot$ are more probable for
the progenitors of PSR 2303+46 and PSR 1518+4904, respectively, with
helium-star masses of $\sim 3$\,M$_\odot$ in each case. 

We use our knowledge of the parameters of DNS progenitors to calculate
the center-of-mass velocities, $V_{\rm CM}$, acquired by the systems
after the explosion (equation 34 in Kalogera 1996), for a specific kick
magnitude. For each observed system, the lowest possible value of $V_{\rm
CM}$ corresponds to the minimum kick magnitude needed for its formation,
and to the progenitor binary with the lowest orbital relative velocity
(widest orbit and least massive helium star). For the two systems in
close orbits (PSR 1913+16 and 1534+12) we find minimum center-of-mass
velocities of 200\,km/s and 225\,km/s, respectively. Since the results of
our synthesis calculations of DNS formation favor low kick magnitudes,
these minimum center-of-mass are also the most probable. We note that if
the age of the system is longer than $\tau_c$, for example is equal to
10\,$\tau_c$, then the center-of-mass velocities are reduced to 100\,km/s
and 185\,km/s, respectively. For the two wider systems, the minimum
values of $V_{\rm CM}$ are considerably lower, $V_{\rm CM}\approx
50$\,km/s.  Measurements of proper motions of the observed DNSs have been
reported in the literature the past few years (Taylor
1993\markcite{Tay93}; Arzoumanian\markcite{Arz95} 1995; 
Frail\markcite{Fra96} 1996; Nice\markcite{N96} 1996). The inferred values
of {\em transverse} velocities for the two close systems, PSR 1913+16 and
1534+12, are $\approx$110\,km/s and $\approx$160\,km/s. For PSR 1518+4904
only an upper limit of 40\,km/s can be set, while for PSR 2303+46 there
is no significant measurement.  We note, however, that these values
depend sensitively on the distance estimates. Moreover, these
measurements include only the transverse component of the space
velocities and the number of systems is too low for any statistically
significant conclusions to be drawn.  Although we should bear in mind
that many uncertainties are involved and the current proper motion data
do not allow any reliable conclusions to be drawn, we note that our
results are consistent with the trend of close systems having higher
center-of-mass velocities. 

\acknowledgments

We are grateful to Dimitrios Psaltis for invaluable discussions and
comments on the manuscript. We thank Ron Webbink for a critical review of
the manuscript, and also Fred Lamb, Doug Swesty, and Icko Iben for
participating in a discussion about kicks at the early stages of this
work. It is also a pleasure to thank Stan Woosley for kindly providing us
with unpublished helium star models. We much appreciate Dale Frail's and
David Nice's efforts to collect and provide us with current measurements
of DNS proper motions, as well as helpful discussions with Paul Harding
and Jim Liebert on the observational prospects of the pulsar companions.
CLF acknowledges support from National Science Foundation under grant AST
92-06738 and VK from National Science Foundation under grant AST92-18074
and the Graduate College of the University of Illinois for a Dissertation
Completion Fellowship.

\newpage

\figcaption{Post-supernova orbital separation, $A$, and eccentricity,
$e$, as a function of the time, $T_{\rm SN}$, elapsed since the
supernova explosion, for PSR 1913+16 and PSR 1534+12. The vertical dotted
lines lie at $T_{\rm SN}=\tau_c$, where $\tau_c$ is the pulsar
characteristic age (Table 1).}

\figcaption{Pre-supernova orbital separation, $A_o$, mass of the exploding
(helium)  star, $M_o$, and ratio, $A_o/R_{\rm He,max}$, of the
pre-SN orbital separation to the maximum radius reached by a helium star
prior to its explosion, as a function of the time, $T_{\rm SN}$, elapsed
since the explosion.  Quantities are plotted for PSR 1913+16 and PSR 1534+12
under the assumptions of circular pre-supernova orbits and symmetric
supernovae. The vertical dotted lines lie at $T_{\rm SN}=\tau_c$, where
$\tau_c$ is the pulsar characteristic age (Table 1).}

\figcaption{Ratio, $A_o(1-e_o)/R_{\rm He,max}$, of the pre-SN
periastron distance to the maximum radius reached by a helium star prior to
its explosion, as a function of the mass, $M_o$, of the helium star, for a
complete range of pre-supernova eccentricities. Plots are shown for PSR
1913+16 and PSR 1534+12, under the assumption of symmetric supernovae
explosions and $T_{\rm SN}=\tau_c$. A minimum helium star mass for
neutron
star formation of 2\,M$_\odot$ has been assumed.}

\figcaption{Limits on the pre-supernova orbital separation, $A_o$, and
mass of the helium star, $M_o$ of (a) PSR 1913+16 and (b) PSR 1534+12,
for three different magnitudes of the kick velocity, $V_k$. The
vertical thick dotted line corresponds to the maximum orbital separation
set by the geometrical constraint;  the thick solid line corresponds to
the minimum helium-star mass that can be accommodated in the orbit; thin
lines correspond to the maximum possible pre-supernova helium-star mass
($\cos^2\phi =0$) and their position depends on the kick magnitude. 
Limits are calculated for circular pre-supernova orbits and for
$T_{\rm SN}=\tau_c$.}

\figcaption{Limits on the pre-supernova orbital separation, $A_o$, and
mass of the helium star, $M_o$ of (a) PSR 1913+16 and (b) PSR 1534+12,
for three different values of the time, $T_{\rm SN}$, elapsed
since the supernova in units of the pulsar characteristic age, $\tau_c$. 
The thick solid line corresponds to the minimum helium-star mass that can
be accommodated in the orbit; thin lines correspond to the limiting cases
of $\cos^2\phi =1$ (vertical lines) and $\cos^2\phi =0$, and depend on
the value of $T_{\rm SN}$. Limits are calculated for circular
pre-supernova orbits and for kick velocity magnitudes (a) $V_k=350$\,km/s
and (b) $V_k=300$\,km/s.}

\figcaption{Limits on the pre-supernova orbital separation, $A_o$, and mass
of the helium star, $M_o$ of (a) PSR 1913+16 and (b) PSR 1534+12, for
eccentric pre-supernova orbits, for $T_{\rm SN}=\tau_c$, and for
different
values of the kick magnitude, $V_k$.}

\end{document}